\documentclass[fleqn,usenatbib]{mnras}

\usepackage{newtxtext,newtxmath}

\usepackage[T1]{fontenc}

\DeclareRobustCommand{\VAN}[3]{#2}
\let\VANthebibliography\thebibliography
\def\thebibliography{\DeclareRobustCommand{\VAN}[3]{##3}\VANthebibliography}

\usepackage{graphicx}	
\usepackage{amsmath}	

\usepackage{xcolor}

\title[GCVD: Globular cluster velocity dispersion distances]{Measuring distances to galaxies with globular cluster velocity dispersions}

\author[Beasley et al.]{
Michael A. Beasley,$^{1,2,3}$\thanks{E-mail: mikey.beasley@gmail.com}
Katja Fahrion,$^{4}$
and Anastasia Gvozdenko, $^{5}$
\\
$^{1}$Centre for Astrophysics and Supercomputing, Swinburne University, John Street, Hawthorn VIC 3122, Australia\\
$^{2}$Instituto de Astrofísica de Canarias, Calle Vía Láctea, E-38206 La Laguna, Spain\\ 
$^{3}$Departamento de Astrofísica, Universidad de La Laguna, E-38206 La Laguna, Spain\\
$^{4}$European Space Agency (ESA), European Space Research and Technology Centre (ESTEC), Keplerlaan 1, 2201 AZ Noordwijk, the
Netherlands.\\
$^{5}$Department of Astrophysics/IMAPP, Radboud University, PO Box 9010,
6500 GL, The Netherlands
}

\date{Accepted XXX. Received YYY; in original form ZZZ}

\pubyear{2023}

\begin{document}
\label{firstpage}
\pagerange{\pageref{firstpage}--\pageref{lastpage}}
\maketitle

\begin{abstract}
Accurate distances are key to obtaining intrinsic properties of astronomical objects such as luminosity or size. Globular clusters (GCs) follow a well-defined relation between their absolute magnitudes and internal stellar velocity dispersions ($\sigma$), offering an independent way to measure distances to their host galaxies via high-resolution spectroscopy. This is reminiscent of the "Faber-Jackson" for elliptical galaxies. However, unlike galaxies, GCs have a very narrow range of mass-to-light ratios and simple star formation histories. Here we show that the GC $M_V - \text{log}_{10}(\sigma)$ relation is linear, whose slope is identical for the Milky Way and M31 GC systems. Based on this, we use 94 Milky Way GCs which have distances from Gaia parallaxes, or  proper-motion dispersion profiles to derive a "GC velocity dispersion" distance (GCVD) to M31, obtaining $(m-M)_0=24.51\pm0.08$ ($d=798\pm28$ kpc), in excellent agreement with independent measurements. Combining data for these two galaxies to create a fiducial relation using 296 GCs with high-quality measurements, we obtain a zeropoint uncertainty ($\pm 0.06$ mag) corresponding to a distance uncertainty of $\sim3\%$.  
We then use GCVD to obtain a distance to the giant elliptical galaxy NGC\,5128 (Centaurus A), finding $(m-M)_0= 27.95\pm0.09$ ($d=3.89\pm0.16$ Mpc). This is in excellent agreement with, and in some cases more precise than, literature estimates from the tip of the red giant branch or surface brightness fluctuations. We apply GCVD to Local Group galaxies with appropriate  data and find good agreement with literature values even in cases with only one GC velocity dispersion measurement.
\end{abstract}




\section{Introduction}

The problem of measuring distances is a fundamental one in astronomy. Without precise distances, intrinsic properties such as physical size and luminosity cannot be determined. As such, a plethora of distance measurement techniques have been developed. Most commonly, astronomical objects are identified which have known intrinsic luminosities and therefore act as standard candles.  These include, but are not limited to: exploding stars in the case of supernovae (SNe)~\citep{Riess1996ApJ}, variable stars such as Cepheids~\citep{Madore1991}, the tip of the red giant branch (TRGB) in evolved stellar populations~\citep{Lee1993},  the turn-over of the globular cluster luminosity function~\citep{HarrisRacine1979}, surface brightness fluctuations (SBF)~\citep{TonrySchneider1988}, late-type galaxies in the Tully--Fisher relation~\citep{Tully1977} and early-type galaxies in the Faber--Jackson relation (\citealt{Minkowski1962}; \citealt{FaberJackson1976}). 

Aside from the important task of understanding the physical properties of relatively nearby galaxies, some of the techniques 
mentioned above also act as "rungs" of distance ladders used to measure the Hubble constant ($H_0$). Indeed, the need for both accurate and precise distances has increasingly come into focus, especially given the apparent discrepancy between measurements of $H_0$ locally, and that predicted from fits to the cosmic microwave background acoustic power spectrum (the so-called "Hubble tension") (e.g., \citealt{DiValentino21}).

In this context an interesting example is the widespread use of the TRGB to anchor distances to nearby SN Ia host galaxies. The TRGB is often a distance measurement of choice due to its potentially high statistical precision and ubiquity in old stellar populations \citep{Lee1993}. However, the TRGB in individual galaxies has been shown to vary by up to 0.2 mag, possibly due to variations in the stellar populations probed in the host galaxy (e.g. \citealt{Barker2004}; \citealt{Mcquinn2019} -- but see \citealt{Madore2023} for a more optimistic view). Therefore, the development of alternative approaches to measuring distances is a worthwhile pursuit.

It has been known for some time that globular clusters (GCs) in the Milky Way follow a relation between their absolute magnitudes and internal stellar velocity dispersions ($\sigma$) (e.g., \citealt{DubathGrillmair1997}). This stems from the Virial theorem; the random motions of stars in GCs increase with mass and, for a given mass-to-light ratio, more massive GCs are more luminous. Therefore,  like the Faber-Jackson relation  for early-type galaxies, the $M_x - \sigma$ relation (where $M_x$ indicates absolute magnitude in any given band) for GCs can be used as a distance indicator. This possibility was first explored by \cite{PaturelGarnier1992}, who measured distances to the Large Magellanic Cloud (LMC) and M31 based on 18 Milky Way GCs.

On deeper consideration, obtaining distances using GC stellar velocity dispersions has a number of advantages over using galaxy velocity dispersions. Unlike galaxies, GCs generally occupy a very narrow range of mass-to-light ratios (e.g., \citealt{BaumgardtHilker2018, Strader2011}), fundamentally limiting the intrinsic scatter in $M_x - \sigma$. In addition, GCs, unlike galaxies,  are generally uniformly old and have relatively simple star formation histories, reducing the potential for systematic deviations from the relation. Finally, GCs are seen in nearly all galaxies irrespective of mass and morphological type, and show very similar physical properties among these galaxies (\citealt{BrodieStrader2006}; \citealt{Beasley2020}). This suggests that distances obtained using GC $M_x - \sigma$ are not restricted to any given galaxy type, and GC velocity dispersions potentially offer the promise of a distance measurement  unbiased by galaxy mass, morphological type or stellar population.

While the measurement of a single velocity dispersion for a GC can, in principle,  yield a distance to the GC (and therefore its host galaxy), perhaps the true power of using GCs comes from the fact that in many galaxies they constitute populous systems. Even low-mass dwarfs can have relatively rich of systems of GCs, while most massive giant ellipticals may have tens of thousands (e.g. \citealt{Harris2013}). As we will show, using an ensemble of GC velocity dispersions in an $M_x - \sigma$ relation can reduce statistical uncertainties on the distance to levels competitive with "primary" distance indicators such as the TRGB and SBF.

In the following, we explore in some detail the potential of "GC velocity dispersion" distances (hereafter, GCVD). In Section~\ref{Data} we introduce and discuss the principal data we use for our analysis. In Section~\ref{Analysis} we construct GC samples for the Milky Way and M31 to show that their $M_V-\sigma$ relations are statistically identical. We use the $M_V-\sigma$ relation for the Milky Way to measure an independent distance to M31, and then combine these samples to make a fiducial  relation. We apply this fiducial relation to determine the distance to the nearby massive elliptical galaxy Centaurus\,A (NGC\,5128) in addition to a number of Local Group galaxies which have available data. In Section~\ref{Systematics} we discuss some of the possible systematic uncertainties of GCVD, and also look at areas for potential improvements in the method. Finally, in Section~\ref{Discussion}, we summarize our work and look at some potential applications of GCVD.

\section{Data}
\label{Data}

\subsection{Milky Way GC Sample}
\label{MilkyWayGCSample}

Due to their relative proximity, the stellar velocity dispersions of Milky Way GCs are generally obtained from velocity measurements of individual stars. This presents a number of observational challenges and has led to a situation where different studies are not always directly comparable.
Given this situation, we decided to draw our entire Milky Way GC sample from \cite{BaumgardtHilker2018} who measured mass-weighted central velocity dispersions and dispersion profiles for 112 GCs in a homogeneous manner. Central velocity dispersions were determined based on N-body fits to individual stellar velocities and surface brightness profiles of the GCs. The stellar velocities were obtained from a variety of instruments including FLAMES, UVES, X-Shooter and FORS2 on VLT, and DEIMOS, HIRES and NIRSPEC on Keck. Please refer to \cite{BaumgardtHilker2018} for the original sources of these data.

From this sample, we removed GCs with very low velocity dispersions ($\sigma < 1.5$ km s$^{-1}$) which have little leverage on overall distance estimates but add to the scatter in the relation (see Section~\ref{afiducialrelation}).
This cut left us with 94 Milky Way GCs. In addition to the velocity dispersion data, we use extinction-corrected absolute $V$-band magnitudes and $M/L_V$ ratios, derived in a homogeneous manner, based on either distances from Gaia DR2/DR3 parallaxes, or proper motion kinematics \citep{Baumgardt2021}, or the literature \citep{Baumgardt2020}.
The final sample of Milky Way GCs has $1.35\leq~M/L_V\leq3.15$ with a mean $M/L_V=1.87$.

\subsection{M31 GC Sample}
\label{M31GCSample}

We base our M31 sample on the catalogue of \cite{Strader2011} who measured central and global dispersions for 200 M31 GCs using high-resolution, integrated light spectroscopy with MMT/Hectoechelle. 
Of this sample, we use 194 GCs which have $M/L_V$ measurements.
We have included an additional 7 outer halo GCs from \cite{Sakari2015}, as well as the  massive, metal-poor M31 GC Ext 8 \citep{Larsen2022}. This leaves us with a total of 202 M31 GCs. 
The original $V$-band magnitudes and extinction values used in \cite{Strader2011} were taken from the catalogues of \cite{Caldwell2009, Caldwell2011}, which are themselves revisions of the catalogues from the "Bologna Group" \citep{Galleti2006}. 
We note that the minimum velocity dispersion for GCs in the M31 sample is 2.1 km s$^{-1}$, which is slightly above our cuts for the Milky Way GCs (see Sec.~\ref{MilkyWayGCSample}). The M31 GCs considered here have $0.27\leq~M/L_V\leq4.05$ with a mean $M/L_V=1.62$.

\subsection{Centaurus A (NGC 5128)}
\label{CenAData}

Stellar velocity dispersion data exists for a good number of GCs in NGC\,5128 from the works of \cite{Taylor2015} (115 GCs) and \cite{Dumont2022} (57 GCs).
\cite{Dumont2022} present global velocity dispersions, foreground extinction-corrected $V$-band absolute magnitudes and $M/L_V$ for 57 luminous GCs in NGC~5128. Since absolute magnitudes are given,  we converted back to GC $V$-band apparent magnitudes using the distance assumed in the paper ($(m-M)_0=27.91$). 
In the \cite{Dumont2022} study, one object (VH81-01) -- the most massive in their sample -- was identified with $M/L_V=7.16^{+1.16}_{-1.0}$.  By comparison, the rest of the sample has mean $M/L_V=2.1\pm0.3$. Therefore, to create a dataset comparable to our fiducial sample in terms of $M/L_V$, we  exclude this cluster leaving us with 56 GCs. We note, however, that including or excluding VH81-01 has a negligible impact on our analysis.

In contrast, \cite{Taylor2015} find a significant number of NGC\,5128 GCs which appear to have elevated $M/L_V$ ratios. For the combined Milky Way + M31 sample, there are no GCs with $M/L_V > 4.05$ whereas there are 37 GCs with $M/L_V$ larger than this value in the \cite{Taylor2015} sample. While the most massive GC-like objects are sometimes seen to have slightly elevated $M/L_V$ -- perhaps indicating that they are in fact stripped nuclei of dwarf galaxies which harbour massive black holes (e.g., \citealt{Ahn2018}) -- those in \citealt{Taylor2015} are not necessarily the most massive and, in some cases, have $M/L_V \gg 10$. It is presently not clear what these objects are, although we note that several papers (e.g., \citealt{Voggel18}, \citealt{Dumont2022}) have questioned the reliability of the M/L from some of the lower S/N data from \citealt{Taylor2015}. In view of the above, we have selected to only use the data of \cite{Dumont2022}.

\section{Analysis}
\label{Analysis}

\subsection{A fiducial relation}
\label{afiducialrelation}

Our first goal is to create a fiducial relation between absolute magnitude and velocity dispersion based on the Milky Way and M31 data. 
The inclusion of the M31 GCs is desirable since it extends the relation to both higher $\sigma$, and also higher metallicity. Both these qualities are useful when applying our methodology to massive galaxies in particular.

To be able to combine the two data sets into a single fiducial relation, we need to demonstrate that both systems obey similar scaling relations. In addition, to be of maximum utility as an extragalactic distance measurement, we wish to use global velocity dispersion measurements (i.e., the dispersion averaged over the entire cluster) rather than central measurements, since the former is what is observable in unresolved GCs.  Moreover, the M31 data needs to be corrected to absolute magnitudes in such a way that they are independent of other distance measurements.

To obtain global velocity dispersions ($\sigma_{\rm gl}$) for the Milky Way data, we have applied  the virial theorem (e.g., \citealt{DubathGrillmair1997}):

\begin{equation}
\sigma_{\rm gl}^2 = \frac{M_{\rm vir}~G}{7.5~r_{\rm hm}},
\label{eq1}
\end{equation}

where $M_{\rm vir}$ is the virial mass -- which we equate to the total mass given in 
\cite{BaumgardtHilker2018} -- and $r_{\rm hm}$ is the half-mass radius, again taken from \cite{BaumgardtHilker2018}. The correction from central to global velocity dispersion goes in the sense that $\sigma_{\rm gl}$ is always lower than the central velocity dispersion by an average of $\sim23\%$. For subsequent corrections, we derive a convenience function in the form of a linear model $\sigma_{\rm gl} = a + b\sigma_{\rm cen}$, where  $\sigma_{\rm cen}$ is the central velocity dispersion.
An unweighted, linear least-squares fit yields $a=0.085\pm0.064$ and $b=0.752\pm0.009$. 
We apply this correction to the Milky Way data and any subsequent data which consists of central velocity dispersions in the absence of aperture effects. The uncertainties in the correction are propagated into the total uncertainty for the corrected velocity dispersions and are of order $\sim0.1$ km s$^{-1}$.

\begin{figure}
	\includegraphics[width=\columnwidth]{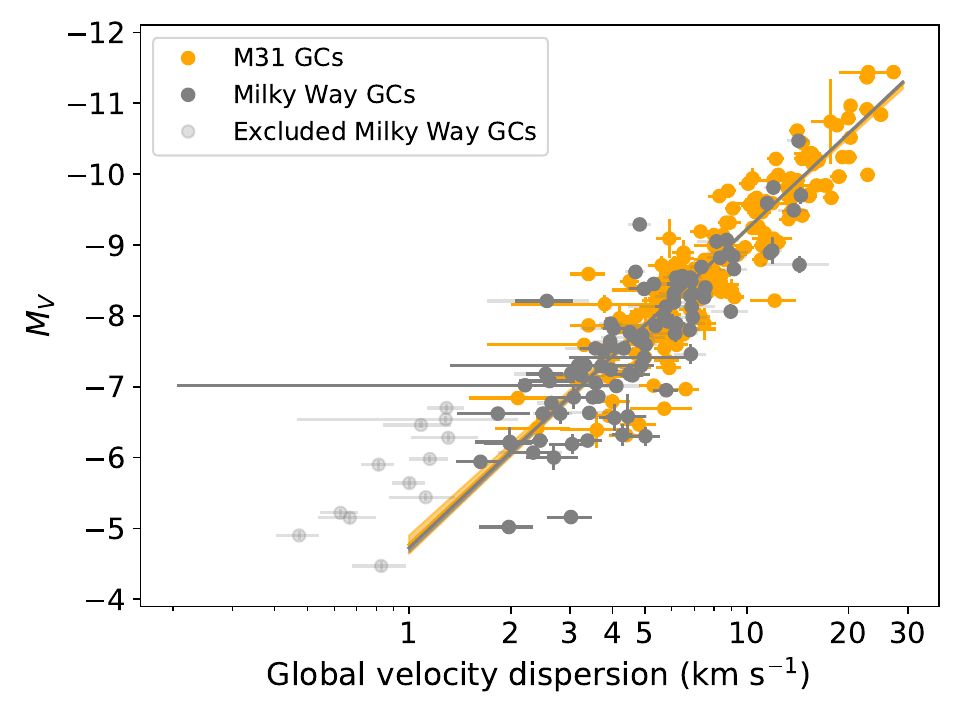}
    \caption{Relation between $M_V$ and velocity dispersion for Milky Way GCs (dark grey) and M31 GCs (orange), the latter converted to absolute magnitudes via GCVD (see text). The lines represent the linear relations for the GCs as determined from MCMC. The shaded bands indicate the 16th, 84th percentile confidence intervals from MCMC. Milky Way GCs excluded from the analysis are shown as faint grey symbols (see Sec.~\ref{MilkyWayGCSample}). Note that the uncertainties on $M_V$ are in many cases smaller than the symbol size.}
    \label{fig1}
\end{figure}

The corrected Milky Way GC data are shown in Fig.\,\ref{fig1}. Inspection of the figure indicates that the data can be well-approximated by a linear model:

\begin{equation}
    M_V = \beta_0 + \beta_1{\rm log_{10}}(\sigma_{\rm{gl}}),
    \label{eq2}
\end{equation}
where $\beta_0$ is the zeropoint and $\beta_1$ the slope of the relation.

After some experimentation, we elected to use an MCMC ensemble sampler (\textsc{emcee}; \citealt{Foreman-Mackey2013}) to determine the most likely values and uncertainties for $\beta_0$ and $\beta_1$. We adopt a Gaussian likelihood function and include uncertainties in both $\sigma_{\rm gl}$ and $M_V$. Our initial guess for the starting parameters and priors for MCMC were determined from ordinary linear least-squares. We chose uniform priors in the range $-10 < \beta_0 < 31$, and $-10 < \beta_1 <0$, with a burn-in of 100. The large range in $\beta_0$ encompasses distances  from the Milky Way out to approximately 100 Mpc. We ran 10,000 chains and the uncertainties on the parameters are given as the 16th and 84th percentiles of the resulting distributions\footnote{Python code for GCVD is available at \url{https://github.com/mikeybeasley/GCVD}}.  
We tested that the MCMC had converged in several ways. We examined output plots of chains to ensure the absence of any obvious drifts or trends. We also tested for a stable output posterior distribution by starting the chains from different initial conditions and different chain lengths.

The results for the Milky Way and M31 are given in Tab. \ref{tab1}. We find that, within the (1-$\sigma$) uncertainties on the parameters, the slopes for the Milky Way and M31 relations are identical. This is to be expected since the GCs in both galaxies have very similar $M/L_V$ ratios (see Sec.~\ref{Data}). While cannot  claim that statistically the two samples are drawn from the same parent population, the $M_V - \text{log}_{10}(\sigma)$ relations of the Milky Way and M31 samples do appear to be indistinguishable given the present uncertainties.

In principle, we can use the zeropoint returned from MCMC for M31 to obtain a distance to this galaxy. However, precision is improved significantly if we fix the slope of the relation to that of the Milky Way, and re-run MCMC. This new zeropoint for M31 is given in Tab. \ref{tab2}, where we see that the precision is improved by a factor of $\sim3.$ 
By fixing the slope, the intercept ($\beta_0$) of M31 in conjunction with the zeropoint of the Milky Way relation then yields the distance to the galaxy. We fix the slope of the M31 GCs\footnote{We fix the slope in the MCMC code by fixing the value of $\beta_1$ in the likelihood function. In principle, we could have also done this by setting the prior on $\beta_1$ to a very narrow range about the fiducial value, but in practice, this proved numerically awkward -- in particular when it came to producing the likelihood contours and confidence intervals.} to $\beta_1=-4.72$, the value from the Milky Way relation, and obtain a zeropoint of $\beta_0=20.019^{+0.029}_{-0.028}$ (see Tab.~\ref{tab2}). 

Since the M31 $V-$band magnitudes are already corrected for foreground reddening, we then convert to a distance modulus via:

\begin{equation}
    (m - M)_0 =  \beta_0 \rm{(galaxy)} - \beta_0 \rm{(fiducial)},
    \label{eq3}
\end{equation}

\noindent where in this case $\beta_0 (\rm{galaxy})$ corresponds to the zeropoint of M31 and $\beta_0 (\rm{fiducial})$ corresponds to the zeropoint of the Milky Way GCs. Here we find, $(m-M)_0$ (M31) = $24.51\pm0.08$ mag, or $d = 798\pm28$ kpc. This value is in excellent agreement with literature estimates (Tab.~\ref{tab2}). The M31 GCs, shifted by this distance modulus, are compared to the Milky Way GCs in Fig. \ref{fig1}.

The figure indicates very similar relations for the Milky Way and M31, the principal difference being that the M31 GCs extend to higher values of $\sigma_{\rm gl}$.
Given the  similarities of the two relations, we combine the Milky Way and M31 samples (with M31 shifted by our distance determination) into a single "fiducial relation" totalling 296 GCs.  Our MCMC determinations for $\beta_0$ and $\beta_1$ for this fiducial relation are  listed in Tab. \ref{tab1}.  We find a zeropoint, $\beta_0=-4.49^{+0.04}_{-0.04}$, which corresponds to a statistical distance uncertainty of $\sim3\%$. We note that the fiducial relation is nearly identical to the Milky Way relation alone, with the advantages of marginally higher precision and a larger coverage in $\sigma_{\rm gl}$. Corner plots showing our MCMC results for the Milky Way, M31 and the fiducial sample are shown in Appendix A.

In principle,  the relation is valid for $1.5 < \sigma_{\rm{gl}} < 27.2$ km s$^{-1}$, which is the range of the fiducial data. However, in practice we note that for $\sigma_{\rm{gl}}< 4$ km s$^{-1}$ the scatter increases significantly. This arises due to the fact that the  $M_V - \sigma_{\rm{gl}}$ relation (i.e., $\sigma_{\rm{gl}}$ on a linear scale) flattens out in such a way that discriminating power for $M_V$ is lost. In addition, we see indications that  the GCs with $\sigma_{\rm{gl}} < 1.5$ km s$^{-1}$ diverge from the fiducial relation. This could be a real effect, or it could be a result of measurement uncertainties of the individual GC stars affecting the velocity dispersion measurement. In view of this, and since GCVD loses predictive power at low velocity dispersions, we recommend that only GCs with $\sigma_{\rm{gl}}\geq 4$ km s$^{-1}$ be used for distance estimations.

\subsection{The effect of metallicity}
\label{Theeffectofmetallicity}

\begin{figure}
	\includegraphics[width=\columnwidth]{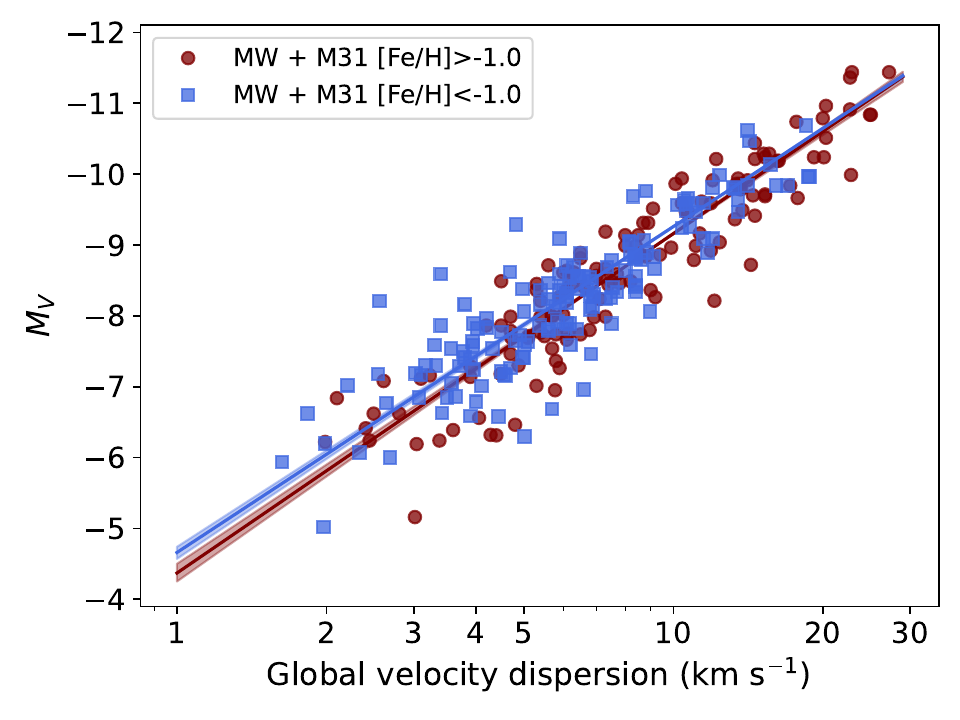}
    \caption{Relation between $M_V$ and velocity dispersion for the metal-poor (blue) and metal-rich (red) combined Milky Way and M31 samples. 
    The lines represent the linear relations and confidence intervals for each of the populations as determined from MCMC. Uncertainties on individual data points are omitted for clarity.}
    \label{fig2}
\end{figure}

One aspect to characterise for GCVD is its possible dependence on metallicity. This is important since GCs in external galaxies show a wide range of metallicities, roughly bracketing $-2.5<$[Fe/H]$<0.0$ (e.g., \citealt{Beasley2019}).
To explore this issue, we split the fiducial sample at [Fe/H] = $-1.0$, creating two sub-samples of metallicity, one metal-poor (151 GCs)  and one metal-rich (135 GCs)( note that the total GCs in the combined sub-samples (286 GCs)  is slightly less that the combined original samples (296 GCs) since not all M31 GCs have good metallicities).  The metal-poor sub-sample has a mean metallicity of [Fe/H]$ = -1.54$, the metal-rich sub-sample has a mean [Fe/H]$ = -0.54$. We then ran MCMC on the sub-samples using the same priors as those we used for the fiducial sample, and the results are given in Tab. \ref{tab1} and shown graphically in Fig.~\ref{fig2}. We find that the slopes of the metal-rich and metal-poor sub-samples are in agreement within the measurement uncertainties on the parameters. However, we do see that the metal-poor sub-sample seems to have a slightly a brighter zeropoint (the difference is $\sim1\sigma$).

In order to isolate the impact of metallicity on the zeropoint, we fix the slopes of the metal-poor and metal-rich sub-samples to that of the full fiducial sample and re-ran MCMC. We find (metal-poor) $\beta_0=-4.55^{+0.06}_{-0.04}$ and (metal-rich) $\beta_0=-4.45^{+0.05}_{-0.04}$. I.e., these zeropoints are consistent within the uncertainties. If we repeat this exercise for GCs with $\sigma_{\rm{gl}}>4$ km s$^{-1}$ -- the GCs which have the most leverage for distances in GCVD -- we again find no significant difference in the zeropoints (metal-poor: $\beta_0=-4.46^{+0.04}_{-0.04}$, metal-rich: $\beta_0=-4.41^{+0.05}_{-0.05}$). Therefore, we find no compelling evidence for a significant dependence of the GCVD zeropoint on metallicity. 

\subsection{A GC velocity dispersion distance to NGC\,5128 (Centaurus\,A)}

We now proceed to use GCVD to obtain a distance to NGC\,5128 (Centaurus\,A). NGC~5128 is a nearby massive elliptical galaxy with an estimated $1450\pm160$ GCs \citep{Hughes2021}. Its cluster system has been studied extensively both photometrically and spectroscopically (e.g., \citealt{Peng2004, Woodley2007, Beasley2008, Hughes2023}).

Having shown that the Milky Way and M31 relations have the same slope within the uncertainties, it is not unreasonable to assume the same slope of the relation for the NGC\,5128 GCs.  Therefore, we fix the slope of the NGC\,5128 GCs to $\beta_1=-4.73$, the value from the fiducial relation, and obtain a zeropoint of $\beta_0=23.458^{+0.062}_{-0.061}$ (see Tab.\ref{tab2}). Note that for convenience, we quote symmetrical uncertainties from hereon and in the following distance determinations, conservatively adopting the larger of the two uncertainties from MCMC. 

\begin{figure}
	\includegraphics[width=\columnwidth]{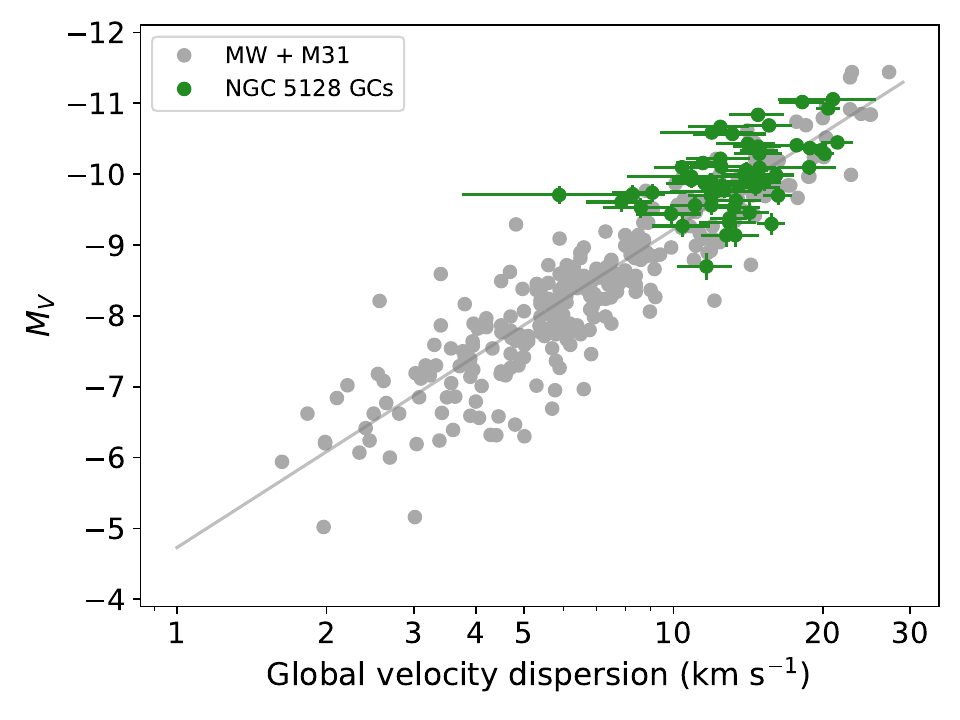}
    \caption{Relation between $M_V$  and velocity dispersion  for the fiducial (Milky Way + M31) GC sample (grey), and the NGC\,5128 sample (green). The NGC\,5128 data have been shifted by our distance determination of $(m-M)_0= 27.94$}
    \label{fig3}
\end{figure}

Since the NGC\,5128 GC $V-$band magnitudes are already corrected for foreground reddening, we then convert to a distance modulus via equation~\ref{eq3}, giving $(m-M)_0$ (NGC\,5128) = $27.95\pm0.09$ mag, or $d = 3.89\pm0.16$ Mpc. The NGC\,5128 GCs, shifted by this distance modulus, are compared to the Milky Way and M31 GCs in Fig. \ref{fig3}.

We derive a total  uncertainty  on the distance modulus for NGC\,5128 as: $\sqrt{\delta\beta^{2}_0{\rm{(fiducial)}} + \delta\beta^2_0\rm{(NGC\,5128)}}$, where $\delta\beta_0\rm{(fiducial)}$ and $\delta\beta_0\rm{(NGC\,5128)}$ are the uncertainties in the fiducial and NGC\,5128 zeropoints, respectively. 

We find that there are 54 distances (dated 7 June 2023) listed for NGC\,5128 in the NASA extragalactic database (NED), obtained using a variety of techniques including TRGB, Cepheids, the GC luminosity function, SBF, Mira variables and SNe Ia. However, it is not obvious which one(s) should be chosen for comparison with our determination. While there may be good  reasons to prefer some methodologies over others, we run the risk of implicit confirmation biases. 
In addition, there is considerable variance even amongst the same distance indicator; to take the example of Cepheid distances alone, the standard deviation of the 4 NED entries is 0.13 mag. For SBF, the standard deviation of the 9 entries is 0.38 mag (after removing one repeat entry).
Therefore, we have simply taken all the distances given in NED and determined the median  and standard deviation of the distribution. From NED we obtain a median distance modulus, $\widetilde{m-M} = 27.82$, with a standard deviation of 0.36. This is in excellent agreement with our determination.

We also explored the possibility of including the NGC~5128 data to further 
augment the fiducial GCVD relation. However, the precision of the relation did not improve with this inclusion, and we therefore leave this issue for when larger samples become available for this galaxy.

\subsection{Local Group Galaxies}

\begin{figure}
	\includegraphics[width=\columnwidth]{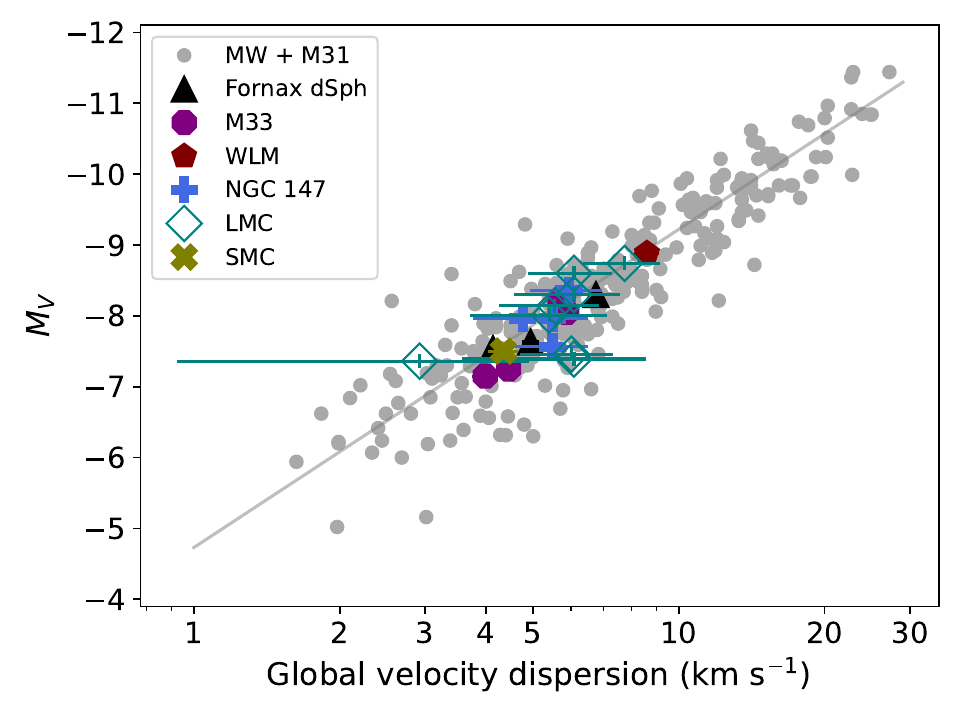}
    \caption{Relation between $M_V$ and velocity dispersion for GCs in Local Group galaxies with data available. The Local Group data have been shifted by the distance moduli obtained from GCVD.}
    \label{fig4}
\end{figure}

Encouraged by the results for M31 and NGC\,5128, we performed a similar analysis for Local Group galaxies which have velocity dispersion data available for some of their GCs. We consider WLM, NGC\,147, M33, NGC\,6822, the Fornax dwarf spheroidal, the Small Magellanic Cloud (SMC) and the Large Magellanic Cloud (LMC). As discussed in Section~\ref{afiducialrelation}, we only select clusters with $\sigma_{\rm{gl}}\geq 4$ km s$^{-1}$. The data for these galaxies is compared to the fiducial relation -- shifted by our GCVD distances -- in Fig.~\ref{fig4}.

\begin{figure*}
	\includegraphics[width=16cm]{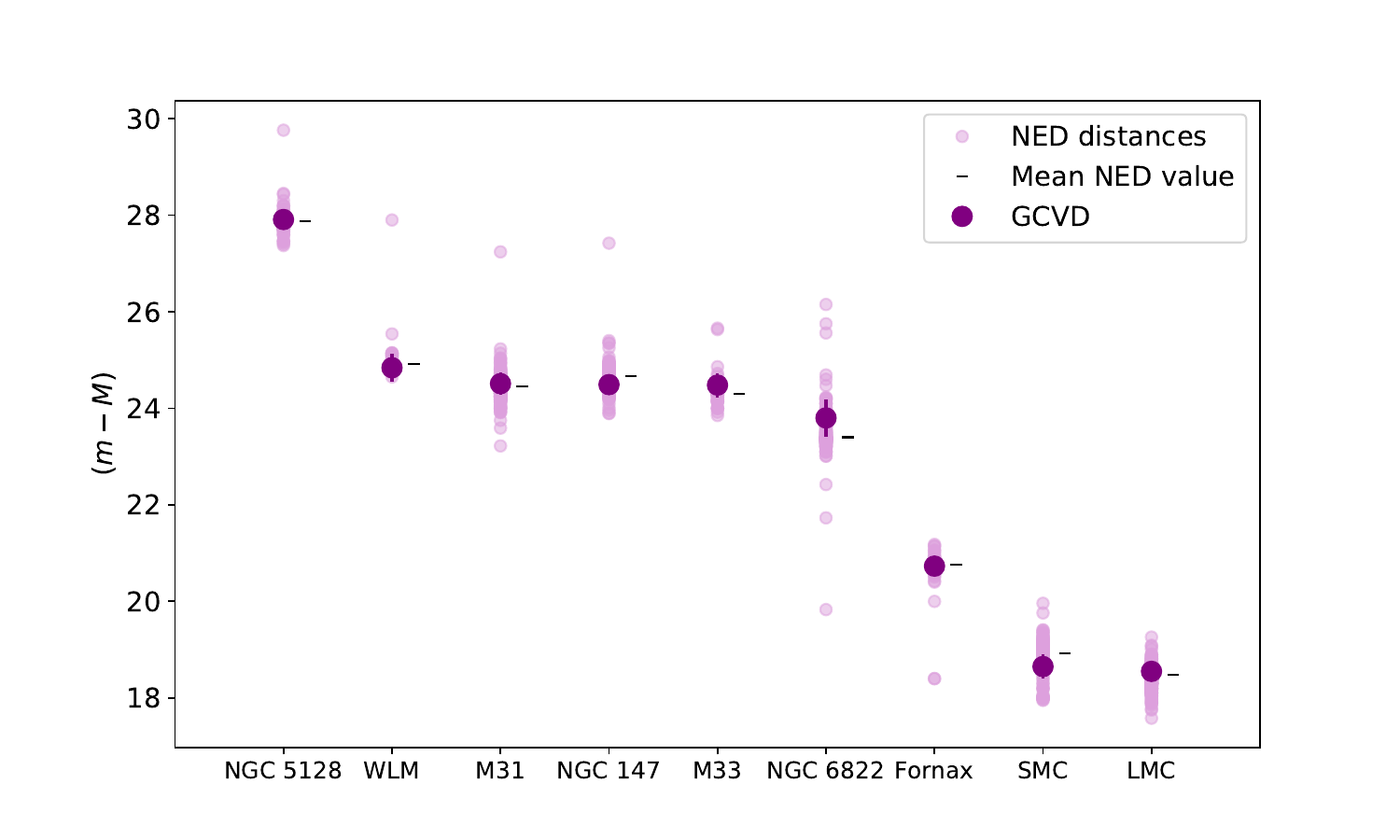}
    \caption{Distance moduli from GCVD (large purple circles with errorbars which are in some cases smaller than the symbol size) compared to those taken from the NED (small pink points). The small black dashed lines indicate the median of the NED values. In all cases, the values of $m-M$ from GCVD agree well with that found from NED.}
    \label{fig5}
\end{figure*}

The results of our fits, and the corresponding NED medians and standard deviations on the distance moduli of individual galaxies are listed in Tab.~\ref{tab2}. and are shown graphically in Fig.~\ref{fig5}. We find, in all cases, good agreement between the GCVD distances and the ensemble of literature values. This is perhaps surprising, given the small number of GCs in some of the galaxies. However, it does suggest that GCVD provides a  robust distance measure across many galaxy types, and that even measurements of $\sigma_{\rm gl}$ for single GCs provide useful constraints on the distance. Further details for the individual galaxies are given below.

\subsubsection{WLM}

The WLM dwarf galaxy is presently only known to host one GC. Its velocity dispersion was taken from \cite{Larsen2014} and $V$-band magnitude ($V=16.06$) from \cite{Sandage1985}. We adopted an extinction of $A_V=0.104$ \citep{Schlafly2011}. The $\sigma$ quoted by \cite{Larsen2014} is neither a central nor global value but is somewhere between the two, modified by aperture effects. We estimate a correction of order $\sim10\%$ downwards based on the corrections listed in \cite{Larsen2002} and our corrections for the Milky Way data.  \cite{Larsen2014} do not quote an uncertainty on their $\sigma$ measurement, and we adopt a conservative uncertainty of 2 km s$^{-1}$. This gives $\sigma_{\rm gl}=8.6\pm 2.0$ km s$^{-1}$. 

\subsubsection{M33}

Global velocity dispersions and $V$-band magnitudes for four GCs (H38, M9, R12, U49) were taken from \cite{Larsen2002}. The extinction-corrected $V$-band magnitudes were taken from \cite{Sarajedini2000}. 

\subsubsection{NGC 147}

\cite{Larsen2018} obtained velocity dispersions for five GCs in this dwarf elliptical satellite of M31. We excluded the GC HII since it has 
$\sigma_{\rm gl} < 4.0$~km s$^{-1}$.
To the four remaining GCs (HIII, PA-1, PA-2, SD7), we apply an aperture correction to their velocity dispersions as mentioned above. We also use the $V$-band magnitudes listed in Table 1 of \cite{Larsen2018}, which originate from \cite{Veljanoski2013}. We assume a foreground extinction $A_V=0.473$ for the GCs \citep{Schlafly2011}. 

\subsubsection{NGC 6822}

\cite{Larsen2018} obtained velocity dispersions for two GCs (SC6, SC7) in this dwarf irregular galaxy. We corrected these dispersions to global values. We used these in conjunction with the $V$-band magnitudes listed in Table 1 of \cite{Larsen2018}, which originate from \cite{Veljanoski2015}. We assume a foreground extinction $A_V=0.646$ for both of the GCs \citep{Schlafly2011}. 

\subsubsection{Fornax dSph}

We took central velocity dispersions, $V$-band magnitudes and extinction corrections for 3 GCs (Fornax 3, 4 \& 5) from \cite{Larsen2012}. The central dispersions were corrected for aperture effects as explained in Section~\ref{afiducialrelation}.

\subsubsection{SMC}

We took central velocity dispersion data for the single old GC NGC 121 from \cite{Dubath1997}. We correct this value to a global dispersion as explained in Section~\ref{afiducialrelation}.
The $V$-band magnitude of the cluster was taken from  \cite{Alcaino1978}, and we used extinction values from \cite{Crowl2001}. 

\subsubsection{LMC}
\label{LMC}

We took central velocity dispersion data for 8 old LMC GCs from \cite{Dubath1997}, corrected to global dispersions as explained in Section~\ref{afiducialrelation}. $V$-band magnitudes were taken from the catalogue of \cite{Bica1996} and we adopted reddenings from \cite{Zaritsky04}. We note that the NED extinctions for these clusters are some 0.15 mag smaller than those of \cite{Zaritsky04}. If we were to adopt the NED values, the LMC modulus would be some 0.15 mag more distant. We did not use additional data from \cite{Song2021} since the two old clusters listed there have very low velocity dispersions which fall below our criteria. 

\section{Systematic effects and improvements}
\label{Systematics}

Given the excellent agreement between GCVD distances and those from the literature, we have reason to believe that GCVD provides both accurate and precise distances. However, this is not to say that GCVD is free from potential systematic problems which may affect its distances.

The most obvious GC property that might influence GCVD is the $M/L_V$ of the GCs observed. A global increase of a factor of 2 in $M/L_V$ will leave the slope of $M_V - \text{log}_{10}(\sigma)$ unaffected, but could potentially shift the zeropoint by 0.75 magnitudes. Reassuringly, such large differences in $M/L_V$ are not seen in entire GC systems. The Milky Way GC system shows a very narrow range of $M/L_V$ which is constant with metallicity \citep{Baumgardt2020}. The M31 GCs do show some indication of a mild decrease of $M/L_V$ with metallicity \citep{Strader2011}, however the similarities between the Milky Way and M31 GCVD relations, and the lack of a dependence of the GCVD zeropoint on metallicity suggests that this variation is not particularly important for GCVD.

 In individual cases, the most luminous GCs ($M_V\gtrapprox-11$, $\sigma\gtrapprox24$ km s$^{-1}$) in populous GC systems do sometimes exhibit high $M/L_V$ ratios, higher than expected for baryonic, old stellar populations (e.g., \citealt{Dumont2022}). This may indicate that they are remnant nuclear star clusters, ultra compact dwarf galaxies (UCDs) with dark matter and/or harbour central massive black holes. Application of our fiducial calibration to such an individual object will systematically overestimate their luminosities, thereby overestimating their distances. In order to address this problem, ideally the target GCs should have their $M/L_V$ determined beforehand. Of course, to determine the GC mass knowledge of the cluster radius is necessary -- a distance dependant quantity. This is also true for the cluster luminosity.  In these cases, an iterative approach using independent distance measurements to filter suitable GCs for GCVD might be necessary. Alternatively, if a reasonable number of GC velocity dispersions are obtained for a galaxy, any extreme outliers with high $\sigma$ may be considered for exclusion from the sample; only observing clusters with \mbox{$\sigma < 20$ km s$^{-1}$} should help mitigate this problem.

A variation in GC ages will also affect $M/L_V$; younger GC ages lead to lower $M/L_V$ for a fixed mass and if systematic -- say in terms of an age metallicity relation (AMR) -- the slope and zeropoint of $M_V - \text{log}_{10}(\sigma)$ may be affected depending upon the magnitude and form of the age variation. For example, at old ages, going from a GC age of 13 to 11 Gyr yields an expected variation of $\sim\,0.3 M_\odot / L_\odot$ (see e.g., \citealt{Baumgardt2020} for a detailed discussion). A systematic age difference of this magnitude for an {\it entire} GC system would be expected to affect the GCVD zeropoint by $\sim0.1$ mag. However, detecting systematic age differences of a few Gyr at old ages is presently very challenging in extragalactic GC systems (e.g., \citealt{Cabreraziri2022}).

Observationally, there are a few ways that we can imagine which may improve the precision  of GCVD. Shifting to infrared passbands for the GC magnitudes would help ameliorate the effects of dust, something which is readily apparent in the extinction corrections for some of the Local Group galaxies we consider (see e.g., Sec.~\ref{LMC}). Indeed some GCs -- such as NGC~2016 in the LMC -- show strong differential reddening across the cluster \citep{Olsen1998} and it is not immediately obvious how to best correct an integrated light spectrum taken from such a system.

In addition, larger samples of velocity dispersions for GCs with good ages and $M/L_V$ would help improve the fiducial calibration,  since the GCs could then be tightly filtered by these quantities while still leaving a sufficient sample size for precise constraints on $M_V - \text{log}_{10}(\sigma)$. For example, in the future one could imagine using larger samples of NGC~5128 GCs as part of a fiducial calibration using an order of magnitude more GCs than that presented here. %

\begin{table*}
	\centering
	\caption{Results for MCMC parameters for Milky Way and M31}
	\label{tab1}
	\begin{tabular}{lccc} 
		\hline
		GC Sample & N. GCs & $\beta_0$ & $\beta_1$ \\
		\hline
		Milky Way & 94 & $-4.49^{+0.05}_{-0.05}$ & $-4.72^{+0.06}_{-0.06}$ \\
		M31 & 202 & $20.011^{+0.09}_{-0.10}$ &  $-4.77^{+0.10}_{-0.09}$\\
        Milky Way + M31 metal-poor & 151 & $-4.60^{+0.08}_{-0.08}$ & $-4.66^{+0.09}_{-0.08}$ \\
        Milky Way + M31 metal-rich & 135 & $-4.37^{+0.12}_{-0.13}$ & $-4.79^{+0.14}_{-0.13}$ \\
        \hline
        Fiducial (Milky Way + M31) & 296 & $-4.49^{+0.04}_{-0.04}$ & $-4.73^{+0.05}_{-0.05}$ \\
        \hline
	\end{tabular}
\end{table*}

\begin{table*}
	\centering
	\caption{Results for MCMC parameters and distances for M31, Centaurus A and Local Group dwarfs}
	\label{tab2}
	\begin{tabular}{lccccc} 
        \hline
		  Galaxy & N. GCs & $\beta_0$  & $(m-M)_0$ & $d$ & NED median (std. dev.) $m-M$ \\ 
         & & (mag) & (mag) & (Mpc) & (mag) \\
		\hline
        M31 & 202 & $20.019\pm0.029$ & $24.509\pm0.076$ & $0.798\pm0.028$ & 24.45 (0.23) \\ 
        Centaurus A & 56 & $23.458\pm0.062$ &$27.948\pm0.089$ & $3.887\pm0.163$ & 27.82 (0.36)\\
        LMC & 8 & $14.072\pm0.185$ &  $18.562\pm0.196$ & $0.053\pm0.005$ & 18.48 (0.16)\\
        M33 & 4 & $19.991\pm0.170$ & $24.481\pm0.182$ & $0.787\pm0.069$ & 24.67 (0.35)\\
        NGC~147 & 4 & $20.001\pm0.253$ &  $24.491\pm0.261$ & $0.791\pm0.101$ & 24.29 (0.36)\\
        Fornax dSph & 3 & $16.242\pm0.187$ &  $20.732\pm0.198$ & $0.140\pm0.013$ & 20.76 (0.41)\\ 
        NGC~6822 & 2 & $19.347\pm0.373$ &  $23.837\pm0.370$ & $0.585\pm0.110$ & 23.40 (0.60)\\  
        SMC & 1 & $14.148\pm0.232$ & $18.638\pm0.241$ & $0.053\pm0.006$ & 18.93 (0.27) \\
        WLM & 1 & $20.353\pm0.270$ & $24.843\pm0.279$ & $0.930\pm0.127$ & 24.92 (0.46) \\
       
        \hline
	\end{tabular}
\end{table*}

\section{Summary and Conclusions}
\label{Discussion}

We have investigated the use of the GC absolute magnitude – internal
stellar velocity dispersion relation (GCVD) in order to obtain independent
distances to galaxies. We first show that the $M_V - \text{log}_{10}(\sigma)$ relation is linear, with a slope which is identical within the uncertainties for the Milky Way and M31 GC systems. We then determine the distance to M31 based on  Milky Way GCVD, finding, $(m-M)_0=24.51\pm0.08$ mag, or $d = 798\pm28$ kpc. This value is in excellent agreement with literature estimates.
We then proceed to construct a fiducial relation using both the Milky Way and M31 GCs. Via MCMC fitting we find a total uncertainty on the zeropoint of this fiducial relation of $\pm0.06$ magnitudes, which corresponds to a distance uncertainty of $\sim3\%$.

As a further example of GCVD, we go on to determine a distance to the giant elliptical galaxy NGC~5128, and obtain $(m-M)_0=27.95\pm0.09$, or $d=3.89\pm0.16$ Mpc. This is in excellent agreement with, and also competitive with, distances derived from Cepheids (e.g., \citealt{Ferrarese2007}; \citealt{Majaess2008}), TRGB (e.g., \citealt{Tully2015}) or SBF (e.g., \citealt{Tonry2001}; \citealt{Ferrarese2007}).

Encouraged by these results we then apply the approach to Local Group galaxies which have velocity dispersion measurements for at least one of their GCs. We find in all cases good agreement between our distances and the ensemble literature determinations from NED. While the uncertainties on our distances are dependent upon the number and precise velocity dispersion distribution of the GCs, we typically obtain distances good to 10\% for these numerically GC-poor systems. 
For example, for the Large Magellanic Cloud we obtain $(m-M)_0=18.56\pm0.20$, while a proposed "consensus distance modulus" for the LMC is $(m-M)_0=18.50$ (e.g., \citealt{Alves2004}). We note that the uncertainties for the Local Group systems can be improved with more and higher quality velocity dispersion measurements for their GCs.

Given that there exist a number of "primary" distance indicators, it is worth asking in what situations GCVD may be applicable and useful.
Firstly, we note in passing that GCVD distances lack fidelity in terms of measuring distances to regions in any single galaxy, this is unlike the situation for, say, TRGB distances. Rather, in principle GCVD gives mean distances to the centre of {\it mass} of the system. This could potentially be exploited to give interesting insights into aspects of the dynamical state of galaxies or galaxy clusters for example.

More generally, accurate distances are not only required to obtain many physical properties of galaxies and their components, but distance measurements are also fundamental for cosmology. For example, measurements of the expansion of the Universe hinge on the cosmic distance ladder that requires high quality distance measurement methods that can be applied in a variety of different galaxies (see \citealt{Freedman2021} and references therein). Being calibrated with GCs in the Milky Way that have direct distance measurements available from Gaia, the GCVD can potentially be used alongside other powerful methods such as TRGB, Cepheids or SNe to measure distances of a large galaxy sample. In particular, as nearly all galaxies with stellar masses $>10^6M_\odot$ have GCs, the GCVD could be used as an unbiased method of measuring distances independent of galaxy type, requiring only ground-based,  high resolution spectroscopy of relatively bright point sources. With a turn-over magnitude of $V = -7.5$ mag, measuring the velocity dispersions of individual, bright GCs can be currently achieved within a few hours on 10m-class telescopes out to several tens of megaparsecs and upcoming 30m-class telescopes and their high-resolution spectrographs will be able to push the applicability of this method even further.

\section*{Acknowledgements}

We thank Michael Hilker for providing uncertainties on the Milky Way central velocity dispersions and Soeren Larsen for useful discussions about the Local Group GC velocity dispersion data. KF acknowledges support through the ESA Research Fellowship Programme. This research has made extensive use of the NASA/IPAC Extragalactic Database (NED), which is funded by the National Aeronautics and Space Administration and operated by the California Institute of Technology. This work made use of Astropy \citep{astropy:2013, astropy:2018, astropy:2022},  NumPy \citep{numpy} and SciPy \citep{scipy}. MAB used ChatGPT to help debug code.

\section*{Data Availability}
The data underlying this article will be shared on reasonable request to the corresponding author.



\bibliographystyle{mnras}
\bibliography{references} 

\begin{thebibliography}{}
\makeatletter
\relax
\def\mn@urlcharsother{\let\do\@makeother \do\$\do\&\do\#\do\^\do\_\do\%\do\~}
\def\mn@doi{\begingroup\mn@urlcharsother \@ifnextchar [ {\mn@doi@} {\mn@doi@[]}}
\def\mn@doi@[#1]#2{\def\@tempa{#1}\ifx\@tempa\@empty \href {http://dx.doi.org/#2} {doi:#2}\else \href {http://dx.doi.org/#2} {#1}\fi \endgroup}
\def\mn@eprint#1#2{\mn@eprint@#1:#2::\@nil}
\def\mn@eprint@arXiv#1{\href {http://arxiv.org/abs/#1} {{\tt arXiv:#1}}}
\def\mn@eprint@dblp#1{\href {http://dblp.uni-trier.de/rec/bibtex/#1.xml} {dblp:#1}}
\def\mn@eprint@#1:#2:#3:#4\@nil{\def\@tempa {#1}\def\@tempb {#2}\def\@tempc {#3}\ifx \@tempc \@empty \let \@tempc \@tempb \let \@tempb \@tempa \fi \ifx \@tempb \@empty \def\@tempb {arXiv}\fi \@ifundefined {mn@eprint@\@tempb}{\@tempb:\@tempc}{\expandafter \expandafter \csname mn@eprint@\@tempb\endcsname \expandafter{\@tempc}}}

\bibitem[\protect\citeauthoryear{{Ahn} et~al.,}{{Ahn} et~al.}{2018}]{Ahn2018}
{Ahn} C.~P.,  et~al., 2018, \mn@doi [\apj] {10.3847/1538-4357/aabc57}, \href {https://ui.adsabs.harvard.edu/abs/2018ApJ...858..102A} {858, 102}

\bibitem[\protect\citeauthoryear{{Alcaino}}{{Alcaino}}{1978}]{Alcaino1978}
{Alcaino} G.,  1978, \aaps, \href {https://ui.adsabs.harvard.edu/abs/1978A&AS...34..431A} {34, 431}

\bibitem[\protect\citeauthoryear{{Alves}}{{Alves}}{2004}]{Alves2004}
{Alves} D.~R.,  2004, \mn@doi [\nar] {10.1016/j.newar.2004.03.001}, \href {https://ui.adsabs.harvard.edu/abs/2004NewAR..48..659A} {48, 659}

\bibitem[\protect\citeauthoryear{{Astropy Collaboration} et~al.,}{{Astropy Collaboration} et~al.}{2013}]{astropy:2013}
{Astropy Collaboration} et~al., 2013, \mn@doi [\aap] {10.1051/0004-6361/201322068}, \href {https://ui.adsabs.harvard.edu/abs/2013A&A...558A..33A} {558, A33}

\bibitem[\protect\citeauthoryear{{Astropy Collaboration} et~al.,}{{Astropy Collaboration} et~al.}{2018}]{astropy:2018}
{Astropy Collaboration} et~al., 2018, \mn@doi [\aj] {10.3847/1538-3881/aabc4f}, \href {https://ui.adsabs.harvard.edu/abs/2018AJ....156..123A} {156, 123}

\bibitem[\protect\citeauthoryear{{Astropy Collaboration} et~al.,}{{Astropy Collaboration} et~al.}{2022}]{astropy:2022}
{Astropy Collaboration} et~al., 2022, \mn@doi [\apj] {10.3847/1538-4357/ac7c74}, \href {https://ui.adsabs.harvard.edu/abs/2022ApJ...935..167A} {935, 167}

\bibitem[\protect\citeauthoryear{{Barker}, {Sarajedini}  \& {Harris}}{{Barker} et~al.}{2004}]{Barker2004}
{Barker} M.~K.,  {Sarajedini} A.,   {Harris} J.,  2004, \mn@doi [\apj] {10.1086/383026}, \href {https://ui.adsabs.harvard.edu/abs/2004ApJ...606..869B} {606, 869}

\bibitem[\protect\citeauthoryear{{Baumgardt} \& {Hilker}}{{Baumgardt} \& {Hilker}}{2018}]{BaumgardtHilker2018}
{Baumgardt} H.,  {Hilker} M.,  2018, \mn@doi [\mnras] {10.1093/mnras/sty1057}, \href {https://ui.adsabs.harvard.edu/abs/2018MNRAS.478.1520B} {478, 1520}

\bibitem[\protect\citeauthoryear{{Baumgardt} \& {Vasiliev}}{{Baumgardt} \& {Vasiliev}}{2021}]{Baumgardt2021}
{Baumgardt} H.,  {Vasiliev} E.,  2021, \mn@doi [\mnras] {10.1093/mnras/stab1474}, \href {https://ui.adsabs.harvard.edu/abs/2021MNRAS.505.5957B} {505, 5957}

\bibitem[\protect\citeauthoryear{{Baumgardt}, {Sollima}  \& {Hilker}}{{Baumgardt} et~al.}{2020}]{Baumgardt2020}
{Baumgardt} H.,  {Sollima} A.,   {Hilker} M.,  2020, \mn@doi [\pasa] {10.1017/pasa.2020.38}, \href {https://ui.adsabs.harvard.edu/abs/2020PASA...37...46B} {37, e046}

\bibitem[\protect\citeauthoryear{{Beasley}}{{Beasley}}{2020}]{Beasley2020}
{Beasley} M.~A.,  2020, in , Reviews in Frontiers of Modern Astrophysics; From Space Debris to Cosmology.
pp 245--277, \mn@doi{10.1007/978-3-030-38509-5_9}

\bibitem[\protect\citeauthoryear{{Beasley}, {Bridges}, {Peng}, {Harris}, {Harris}, {Forbes}  \& {Mackie}}{{Beasley} et~al.}{2008}]{Beasley2008}
{Beasley} M.~A.,  {Bridges} T.,  {Peng} E.,  {Harris} W.~E.,  {Harris} G. L.~H.,  {Forbes} D.~A.,   {Mackie} G.,  2008, \mn@doi [\mnras] {10.1111/j.1365-2966.2008.13123.x}, \href {https://ui.adsabs.harvard.edu/abs/2008MNRAS.386.1443B} {386, 1443}

\bibitem[\protect\citeauthoryear{{Beasley}, {Leaman}, {Gallart}, {Larsen}, {Battaglia}, {Monelli}  \& {Pedreros}}{{Beasley} et~al.}{2019}]{Beasley2019}
{Beasley} M.~A.,  {Leaman} R.,  {Gallart} C.,  {Larsen} S.~S.,  {Battaglia} G.,  {Monelli} M.,   {Pedreros} M.~H.,  2019, \mn@doi [\mnras] {10.1093/mnras/stz1349}, \href {https://ui.adsabs.harvard.edu/abs/2019MNRAS.487.1986B} {487, 1986}

\bibitem[\protect\citeauthoryear{{Bica}, {Claria}, {Dottori}, {Santos}  \& {Piatti}}{{Bica} et~al.}{1996}]{Bica1996}
{Bica} E.,  {Claria} J.~J.,  {Dottori} H.,  {Santos} J.~F.~C. J.,   {Piatti} A.~E.,  1996, \mn@doi [\apjs] {10.1086/192251}, \href {https://ui.adsabs.harvard.edu/abs/1996ApJS..102...57B} {102, 57}

\bibitem[\protect\citeauthoryear{{Brodie} \& {Strader}}{{Brodie} \& {Strader}}{2006}]{BrodieStrader2006}
{Brodie} J.~P.,  {Strader} J.,  2006, \mn@doi [\araa] {10.1146/annurev.astro.44.051905.092441}, \href {https://ui.adsabs.harvard.edu/abs/2006ARA&A..44..193B} {44, 193}

\bibitem[\protect\citeauthoryear{{Cabrera-Ziri} \& {Conroy}}{{Cabrera-Ziri} \& {Conroy}}{2022}]{Cabreraziri2022}
{Cabrera-Ziri} I.,  {Conroy} C.,  2022, \mn@doi [\mnras] {10.1093/mnras/stac012}, \href {https://ui.adsabs.harvard.edu/abs/2022MNRAS.511..341C} {511, 341}

\bibitem[\protect\citeauthoryear{{Caldwell}, {Harding}, {Morrison}, {Rose}, {Schiavon}  \& {Kriessler}}{{Caldwell} et~al.}{2009}]{Caldwell2009}
{Caldwell} N.,  {Harding} P.,  {Morrison} H.,  {Rose} J.~A.,  {Schiavon} R.,   {Kriessler} J.,  2009, \mn@doi [\aj] {10.1088/0004-6256/137/1/94}, \href {https://ui.adsabs.harvard.edu/abs/2009AJ....137...94C} {137, 94}

\bibitem[\protect\citeauthoryear{{Caldwell}, {Schiavon}, {Morrison}, {Rose}  \& {Harding}}{{Caldwell} et~al.}{2011}]{Caldwell2011}
{Caldwell} N.,  {Schiavon} R.,  {Morrison} H.,  {Rose} J.~A.,   {Harding} P.,  2011, \mn@doi [\aj] {10.1088/0004-6256/141/2/61}, \href {https://ui.adsabs.harvard.edu/abs/2011AJ....141...61C} {141, 61}

\bibitem[\protect\citeauthoryear{{Crowl}, {Sarajedini}, {Piatti}, {Geisler}, {Bica}, {Clari{\'a}}  \& {Santos}}{{Crowl} et~al.}{2001}]{Crowl2001}
{Crowl} H.~H.,  {Sarajedini} A.,  {Piatti} A.~E.,  {Geisler} D.,  {Bica} E.,  {Clari{\'a}} J.~J.,   {Santos} Jo{\~a}o F.~C. J.,  2001, \mn@doi [\aj] {10.1086/321128}, \href {https://ui.adsabs.harvard.edu/abs/2001AJ....122..220C} {122, 220}

\bibitem[\protect\citeauthoryear{{Di Valentino} et~al.,}{{Di Valentino} et~al.}{2021}]{DiValentino21}
{Di Valentino} E.,  et~al., 2021, \mn@doi [Classical and Quantum Gravity] {10.1088/1361-6382/ac086d}, \href {https://ui.adsabs.harvard.edu/abs/2021CQGra..38o3001D} {38, 153001}

\bibitem[\protect\citeauthoryear{{Dubath} \& {Grillmair}}{{Dubath} \& {Grillmair}}{1997}]{DubathGrillmair1997}
{Dubath} P.,  {Grillmair} C.~J.,  1997, \mn@doi [\aap] {10.48550/arXiv.astro-ph/9610194}, \href {https://ui.adsabs.harvard.edu/abs/1997A&A...321..379D} {321, 379}

\bibitem[\protect\citeauthoryear{{Dubath}, {Meylan}  \& {Mayor}}{{Dubath} et~al.}{1997}]{Dubath1997}
{Dubath} P.,  {Meylan} G.,   {Mayor} M.,  1997, \aap, \href {https://ui.adsabs.harvard.edu/abs/1997A&A...324..505D} {324, 505}

\bibitem[\protect\citeauthoryear{{Dumont} et~al.,}{{Dumont} et~al.}{2022}]{Dumont2022}
{Dumont} A.,  et~al., 2022, \mn@doi [\apj] {10.3847/1538-4357/ac551c}, \href {https://ui.adsabs.harvard.edu/abs/2022ApJ...929..147D} {929, 147}

\bibitem[\protect\citeauthoryear{{Faber} \& {Jackson}}{{Faber} \& {Jackson}}{1976}]{FaberJackson1976}
{Faber} S.~M.,  {Jackson} R.~E.,  1976, \mn@doi [\apj] {10.1086/154215}, \href {https://ui.adsabs.harvard.edu/abs/1976ApJ...204..668F} {204, 668}

\bibitem[\protect\citeauthoryear{{Ferrarese}, {Mould}, {Stetson}, {Tonry}, {Blakeslee}  \& {Ajhar}}{{Ferrarese} et~al.}{2007}]{Ferrarese2007}
{Ferrarese} L.,  {Mould} J.~R.,  {Stetson} P.~B.,  {Tonry} J.~L.,  {Blakeslee} J.~P.,   {Ajhar} E.~A.,  2007, \mn@doi [\apj] {10.1086/506612}, \href {https://ui.adsabs.harvard.edu/abs/2007ApJ...654..186F} {654, 186}

\bibitem[\protect\citeauthoryear{{Foreman-Mackey}, {Hogg}, {Lang}  \& {Goodman}}{{Foreman-Mackey} et~al.}{2013}]{Foreman-Mackey2013}
{Foreman-Mackey} D.,  {Hogg} D.~W.,  {Lang} D.,   {Goodman} J.,  2013, \mn@doi [\pasp] {10.1086/670067}, \href {https://ui.adsabs.harvard.edu/abs/2013PASP..125..306F} {125, 306}

\bibitem[\protect\citeauthoryear{{Freedman}}{{Freedman}}{2021}]{Freedman2021}
{Freedman} W.~L.,  2021, \mn@doi [\apj] {10.3847/1538-4357/ac0e95}, \href {https://ui.adsabs.harvard.edu/abs/2021ApJ...919...16F} {919, 16}

\bibitem[\protect\citeauthoryear{{Galleti}, {Federici}, {Bellazzini}, {Buzzoni}  \& {Fusi Pecci}}{{Galleti} et~al.}{2006}]{Galleti2006}
{Galleti} S.,  {Federici} L.,  {Bellazzini} M.,  {Buzzoni} A.,   {Fusi Pecci} F.,  2006, \mn@doi [\aap] {10.1051/0004-6361:20065309}, \href {https://ui.adsabs.harvard.edu/abs/2006A&A...456..985G} {456, 985}

\bibitem[\protect\citeauthoryear{{Harris} \& {Racine}}{{Harris} \& {Racine}}{1979}]{HarrisRacine1979}
{Harris} W.~E.,  {Racine} R.,  1979, \mn@doi [\araa] {10.1146/annurev.aa.17.090179.001325}, \href {https://ui.adsabs.harvard.edu/abs/1979ARA&A..17..241H} {17, 241}

\bibitem[\protect\citeauthoryear{{Harris}, {Harris}  \& {Alessi}}{{Harris} et~al.}{2013}]{Harris2013}
{Harris} W.~E.,  {Harris} G. L.~H.,   {Alessi} M.,  2013, \mn@doi [\apj] {10.1088/0004-637X/772/2/82}, \href {https://ui.adsabs.harvard.edu/abs/2013ApJ...772...82H} {772, 82}

\bibitem[\protect\citeauthoryear{Harris et~al.,}{Harris et~al.}{2020}]{numpy}
Harris C.~R.,  et~al., 2020, \mn@doi [Nature] {10.1038/s41586-020-2649-2}, 585, 357

\bibitem[\protect\citeauthoryear{{Hughes} et~al.,}{{Hughes} et~al.}{2021}]{Hughes2021}
{Hughes} A.~K.,  et~al., 2021, \mn@doi [\apj] {10.3847/1538-4357/abf63c}, \href {https://ui.adsabs.harvard.edu/abs/2021ApJ...914...16H} {914, 16}

\bibitem[\protect\citeauthoryear{{Hughes} et~al.,}{{Hughes} et~al.}{2023}]{Hughes2023}
{Hughes} A.~K.,  et~al., 2023, \mn@doi [\apj] {10.3847/1538-4357/acbf43}, \href {https://ui.adsabs.harvard.edu/abs/2023ApJ...947...34H} {947, 34}

\bibitem[\protect\citeauthoryear{{Larsen}, {Brodie}, {Sarajedini}  \& {Huchra}}{{Larsen} et~al.}{2002}]{Larsen2002}
{Larsen} S.~S.,  {Brodie} J.~P.,  {Sarajedini} A.,   {Huchra} J.~P.,  2002, \mn@doi [\aj] {10.1086/344110}, \href {https://ui.adsabs.harvard.edu/abs/2002AJ....124.2615L} {124, 2615}

\bibitem[\protect\citeauthoryear{{Larsen}, {Brodie}  \& {Strader}}{{Larsen} et~al.}{2012}]{Larsen2012}
{Larsen} S.~S.,  {Brodie} J.~P.,   {Strader} J.,  2012, \mn@doi [\aap] {10.1051/0004-6361/201219895}, \href {https://ui.adsabs.harvard.edu/abs/2012A&A...546A..53L} {546, A53}

\bibitem[\protect\citeauthoryear{{Larsen}, {Brodie}, {Forbes}  \& {Strader}}{{Larsen} et~al.}{2014}]{Larsen2014}
{Larsen} S.~S.,  {Brodie} J.~P.,  {Forbes} D.~A.,   {Strader} J.,  2014, \mn@doi [\aap] {10.1051/0004-6361/201322672}, \href {https://ui.adsabs.harvard.edu/abs/2014A&A...565A..98L} {565, A98}

\bibitem[\protect\citeauthoryear{{Larsen}, {Brodie}, {Wasserman}  \& {Strader}}{{Larsen} et~al.}{2018}]{Larsen2018}
{Larsen} S.~S.,  {Brodie} J.~P.,  {Wasserman} A.,   {Strader} J.,  2018, \mn@doi [\aap] {10.1051/0004-6361/201731909}, \href {https://ui.adsabs.harvard.edu/abs/2018A&A...613A..56L} {613, A56}

\bibitem[\protect\citeauthoryear{{Larsen}, {Eitner}, {Magg}, {Bergemann}, {Moltzer}, {Brodie}, {Romanowsky}  \& {Strader}}{{Larsen} et~al.}{2022}]{Larsen2022}
{Larsen} S.~S.,  {Eitner} P.,  {Magg} E.,  {Bergemann} M.,  {Moltzer} C.~A.~S.,  {Brodie} J.~P.,  {Romanowsky} A.~J.,   {Strader} J.,  2022, \mn@doi [\aap] {10.1051/0004-6361/202142243}, \href {https://ui.adsabs.harvard.edu/abs/2022A&A...660A..88L} {660, A88}

\bibitem[\protect\citeauthoryear{{Lee}, {Freedman}  \& {Madore}}{{Lee} et~al.}{1993}]{Lee1993}
{Lee} M.~G.,  {Freedman} W.~L.,   {Madore} B.~F.,  1993, \mn@doi [\apj] {10.1086/173334}, \href {https://ui.adsabs.harvard.edu/abs/1993ApJ...417..553L} {417, 553}

\bibitem[\protect\citeauthoryear{{Madore} \& {Freedman}}{{Madore} \& {Freedman}}{1991}]{Madore1991}
{Madore} B.~F.,  {Freedman} W.~L.,  1991, \mn@doi [\pasp] {10.1086/132911}, \href {https://ui.adsabs.harvard.edu/abs/1991PASP..103..933M} {103, 933}

\bibitem[\protect\citeauthoryear{{Madore}, {Freedman}, {Owens}  \& {Jang}}{{Madore} et~al.}{2023}]{Madore2023}
{Madore} B.~F.,  {Freedman} W.~L.,  {Owens} K.~A.,   {Jang} I.~S.,  2023, \mn@doi [\aj] {10.3847/1538-3881/acd3f3}, \href {https://ui.adsabs.harvard.edu/abs/2023AJ....166....2M} {166, 2}

\bibitem[\protect\citeauthoryear{{Majaess}, {Turner}  \& {Lane}}{{Majaess} et~al.}{2008}]{Majaess2008}
{Majaess} D.~J.,  {Turner} D.~G.,   {Lane} D.~J.,  2008, \mn@doi [\mnras] {10.1111/j.1365-2966.2008.13834.x}, \href {https://ui.adsabs.harvard.edu/abs/2008MNRAS.390.1539M} {390, 1539}

\bibitem[\protect\citeauthoryear{{McQuinn}, {Boyer}, {Skillman}  \& {Dolphin}}{{McQuinn} et~al.}{2019}]{Mcquinn2019}
{McQuinn} K. B.~W.,  {Boyer} M.,  {Skillman} E.~D.,   {Dolphin} A.~E.,  2019, \mn@doi [\apj] {10.3847/1538-4357/ab2627}, \href {https://ui.adsabs.harvard.edu/abs/2019ApJ...880...63M} {880, 63}

\bibitem[\protect\citeauthoryear{{Minkowski}}{{Minkowski}}{1962}]{Minkowski1962}
{Minkowski} R.,  1962, in {McVittie} G.~C.,  ed., ~ Vol. 15, Problems of Extra-Galactic Research. p.~112

\bibitem[\protect\citeauthoryear{{Olsen}, {Hodge}, {Mateo}, {Olszewski}, {Schommer}, {Suntzeff}  \& {Walker}}{{Olsen} et~al.}{1998}]{Olsen1998}
{Olsen} K.~A.~G.,  {Hodge} P.~W.,  {Mateo} M.,  {Olszewski} E.~W.,  {Schommer} R.~A.,  {Suntzeff} N.~B.,   {Walker} A.~R.,  1998, \mn@doi [\mnras] {10.1046/j.1365-8711.1998.01860.x}, \href {https://ui.adsabs.harvard.edu/abs/1998MNRAS.300..665O} {300, 665}

\bibitem[\protect\citeauthoryear{{Paturel} \& {Garnier}}{{Paturel} \& {Garnier}}{1992}]{PaturelGarnier1992}
{Paturel} G.,  {Garnier} R.,  1992, \aap, \href {https://ui.adsabs.harvard.edu/abs/1992A&A...254...93P} {254, 93}

\bibitem[\protect\citeauthoryear{{Peng}, {Ford}  \& {Freeman}}{{Peng} et~al.}{2004}]{Peng2004}
{Peng} E.~W.,  {Ford} H.~C.,   {Freeman} K.~C.,  2004, \mn@doi [\apj] {10.1086/381236}, \href {https://ui.adsabs.harvard.edu/abs/2004ApJ...602..705P} {602, 705}

\bibitem[\protect\citeauthoryear{{Riess}, {Press}  \& {Kirshner}}{{Riess} et~al.}{1996}]{Riess1996ApJ}
{Riess} A.~G.,  {Press} W.~H.,   {Kirshner} R.~P.,  1996, \mn@doi [\apj] {10.1086/178129}, \href {https://ui.adsabs.harvard.edu/abs/1996ApJ...473...88R} {473, 88}

\bibitem[\protect\citeauthoryear{{Sakari}, {Venn}, {Mackey}, {Shetrone}, {Dotter}, {Ferguson}  \& {Huxor}}{{Sakari} et~al.}{2015}]{Sakari2015}
{Sakari} C.~M.,  {Venn} K.~A.,  {Mackey} D.,  {Shetrone} M.~D.,  {Dotter} A.,  {Ferguson} A. M.~N.,   {Huxor} A.,  2015, \mn@doi [\mnras] {10.1093/mnras/stv020}, \href {https://ui.adsabs.harvard.edu/abs/2015MNRAS.448.1314S} {448, 1314}

\bibitem[\protect\citeauthoryear{{Sandage} \& {Carlson}}{{Sandage} \& {Carlson}}{1985}]{Sandage1985}
{Sandage} A.,  {Carlson} G.,  1985, \mn@doi [\aj] {10.1086/113856}, \href {https://ui.adsabs.harvard.edu/abs/1985AJ.....90.1464S} {90, 1464}

\bibitem[\protect\citeauthoryear{{Sarajedini}, {Geisler}, {Schommer}  \& {Harding}}{{Sarajedini} et~al.}{2000}]{Sarajedini2000}
{Sarajedini} A.,  {Geisler} D.,  {Schommer} R.,   {Harding} P.,  2000, \mn@doi [\aj] {10.1086/316807}, \href {https://ui.adsabs.harvard.edu/abs/2000AJ....120.2437S} {120, 2437}

\bibitem[\protect\citeauthoryear{{Schlafly} \& {Finkbeiner}}{{Schlafly} \& {Finkbeiner}}{2011}]{Schlafly2011}
{Schlafly} E.~F.,  {Finkbeiner} D.~P.,  2011, \mn@doi [\apj] {10.1088/0004-637X/737/2/103}, \href {https://ui.adsabs.harvard.edu/abs/2011ApJ...737..103S} {737, 103}

\bibitem[\protect\citeauthoryear{{Song}, {Mateo}, {Bailey}, {Walker}, {Roederer}, {Olszewski}, {Reiter}  \& {Kremin}}{{Song} et~al.}{2021}]{Song2021}
{Song} Y.-Y.,  {Mateo} M.,  {Bailey} J.~I.,  {Walker} M.~G.,  {Roederer} I.~U.,  {Olszewski} E.~W.,  {Reiter} M.,   {Kremin} A.,  2021, \mn@doi [\mnras] {10.1093/mnras/stab1065}, \href {https://ui.adsabs.harvard.edu/abs/2021MNRAS.504.4160S} {504, 4160}

\bibitem[\protect\citeauthoryear{{Strader}, {Caldwell}  \& {Seth}}{{Strader} et~al.}{2011}]{Strader2011}
{Strader} J.,  {Caldwell} N.,   {Seth} A.~C.,  2011, \mn@doi [\aj] {10.1088/0004-6256/142/1/8}, \href {https://ui.adsabs.harvard.edu/abs/2011AJ....142....8S} {142, 8}

\bibitem[\protect\citeauthoryear{{Taylor}, {Puzia}, {Gomez}  \& {Woodley}}{{Taylor} et~al.}{2015}]{Taylor2015}
{Taylor} M.~A.,  {Puzia} T.~H.,  {Gomez} M.,   {Woodley} K.~A.,  2015, \mn@doi [\apj] {10.1088/0004-637X/805/1/65}, \href {https://ui.adsabs.harvard.edu/abs/2015ApJ...805...65T} {805, 65}

\bibitem[\protect\citeauthoryear{{Tonry} \& {Schneider}}{{Tonry} \& {Schneider}}{1988}]{TonrySchneider1988}
{Tonry} J.,  {Schneider} D.~P.,  1988, \mn@doi [\aj] {10.1086/114847}, \href {https://ui.adsabs.harvard.edu/abs/1988AJ.....96..807T} {96, 807}

\bibitem[\protect\citeauthoryear{{Tonry}, {Dressler}, {Blakeslee}, {Ajhar}, {Fletcher}, {Luppino}, {Metzger}  \& {Moore}}{{Tonry} et~al.}{2001}]{Tonry2001}
{Tonry} J.~L.,  {Dressler} A.,  {Blakeslee} J.~P.,  {Ajhar} E.~A.,  {Fletcher} A.~B.,  {Luppino} G.~A.,  {Metzger} M.~R.,   {Moore} C.~B.,  2001, \mn@doi [\apj] {10.1086/318301}, \href {https://ui.adsabs.harvard.edu/abs/2001ApJ...546..681T} {546, 681}

\bibitem[\protect\citeauthoryear{{Tully} \& {Fisher}}{{Tully} \& {Fisher}}{1977}]{Tully1977}
{Tully} R.~B.,  {Fisher} J.~R.,  1977, \aap, \href {https://ui.adsabs.harvard.edu/abs/1977A&A....54..661T} {54, 661}

\bibitem[\protect\citeauthoryear{{Tully}, {Libeskind}, {Karachentsev}, {Karachentseva}, {Rizzi}  \& {Shaya}}{{Tully} et~al.}{2015}]{Tully2015}
{Tully} R.~B.,  {Libeskind} N.~I.,  {Karachentsev} I.~D.,  {Karachentseva} V.~E.,  {Rizzi} L.,   {Shaya} E.~J.,  2015, \mn@doi [\apjl] {10.1088/2041-8205/802/2/L25}, \href {https://ui.adsabs.harvard.edu/abs/2015ApJ...802L..25T} {802, L25}

\bibitem[\protect\citeauthoryear{{Veljanoski} et~al.,}{{Veljanoski} et~al.}{2013}]{Veljanoski2013}
{Veljanoski} J.,  et~al., 2013, \mn@doi [\apjl] {10.1088/2041-8205/768/2/L33}, \href {https://ui.adsabs.harvard.edu/abs/2013ApJ...768L..33V} {768, L33}

\bibitem[\protect\citeauthoryear{{Veljanoski} et~al.,}{{Veljanoski} et~al.}{2015}]{Veljanoski2015}
{Veljanoski} J.,  et~al., 2015, \mn@doi [\mnras] {10.1093/mnras/stv1259}, \href {https://ui.adsabs.harvard.edu/abs/2015MNRAS.452..320V} {452, 320}

\bibitem[\protect\citeauthoryear{Virtanen et~al.,}{Virtanen et~al.}{2020}]{scipy}
Virtanen P.,  et~al., 2020, \mn@doi [Nature Methods] {10.1038/s41592-019-0686-2}, \href {https://rdcu.be/b08Wh} {17, 261}

\bibitem[\protect\citeauthoryear{{Voggel} et~al.,}{{Voggel} et~al.}{2018}]{Voggel18}
{Voggel} K.~T.,  et~al., 2018, \mn@doi [\apj] {10.3847/1538-4357/aabae5}, \href {https://ui.adsabs.harvard.edu/abs/2018ApJ...858...20V} {858, 20}

\bibitem[\protect\citeauthoryear{{Woodley}, {Harris}, {Beasley}, {Peng}, {Bridges}, {Forbes}  \& {Harris}}{{Woodley} et~al.}{2007}]{Woodley2007}
{Woodley} K.~A.,  {Harris} W.~E.,  {Beasley} M.~A.,  {Peng} E.~W.,  {Bridges} T.~J.,  {Forbes} D.~A.,   {Harris} G. L.~H.,  2007, \mn@doi [\aj] {10.1086/518788}, \href {https://ui.adsabs.harvard.edu/abs/2007AJ....134..494W} {134, 494}

\bibitem[\protect\citeauthoryear{{Zaritsky}, {Harris}, {Thompson}  \& {Grebel}}{{Zaritsky} et~al.}{2004}]{Zaritsky04}
{Zaritsky} D.,  {Harris} J.,  {Thompson} I.~B.,   {Grebel} E.~K.,  2004, \mn@doi [\aj] {10.1086/423910}, \href {https://ui.adsabs.harvard.edu/abs/2004AJ....128.1606Z} {128, 1606}

\makeatother
\end{thebibliography}




\appendix

\section{Corner plots from MCMC}

Here we show the parameter correlations for $\beta_0$ (intercept) and $\beta_1$ (slope) from MCMC for the Milky Way, M31 and fiducial (Milky Way + M31) samples. 

\begin{figure*}
	\includegraphics[width=16cm]{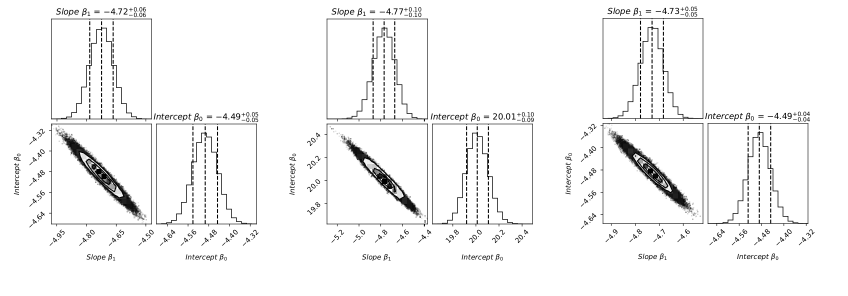}
    \caption{Corner plots for the Milky Way sample (left), the M31 sample (middle -- uncorrected for distance) and the combined Milky Way + M31 (fiducial) sample (right). In these cases, the slope has been left as a free parameter. }
    \label{corners}
\end{figure*}


\bsp	
\label{lastpage}
\end{document}